# Dielectric Models of Normal and Diseased Human Trabecular Bones at Microwave Frequencies

**Bilal Amin[1,2,3], Amra Mehboob[1,2], Eoghan Dunne[1,2,3], MEMBER, IEEE, Ana González-Suárez[1,2,3], Finn Krewer[1,2], Aoife Lowery[4], Martin O'Halloran[1,2,3], Adnan Elahi[1,2], MEMBER, IEEE**

[1]Translational Medical Device Laboratory, University of Galway, H91 TK33 Galway, Ireland
[2]Electrical and Electronic Engineering, University of Galway, H91 TK33 Galway, Ireland
[3]School of Medicine, University of Galway, Galway H91 TK33, Ireland
[4]Discipline of Surgery, Lambe Institute for Translational Research, University of Galway, Galway H91 TK33, Ireland

CORRESPONDING AUTHOR: Bilal Amin (e-mail: bilal.amin@universityofgalway.ie).

**ABSTRACT** The dielectric properties of biological tissues are used to investigate the interaction of electromagnetic (EM) fields with the human body. These properties are integrated into EM numerical simulations, such as the finite difference time domain (FDTD) method, using dielectric models such as Cole-Cole and Debye models. These computational tools are employed to study the propagation of EM waves within the human body. This paper has developed the one-pole and two-pole Cole-Cole and Debye models for the dielectric properties of normal and diseased human trabecular bone tissue at microwave frequencies (0.5 – 8.5 GHz). The dielectric properties of normal trabecular and cortical bone were derived based on the literature, while those of diseased human trabecular bones (osteoporotic and osteoarthritis) were acquired from a previous study conducted by the authors, which characterised the properties within the 0.5 – 8.5 GHz frequency range. This study has employed a weighted least squares method to optimise the one-pole and two-pole dielectric models' parameters. The accuracy of the developed dielectric models is validated by comparing the simulated results with the experimental data already published by the authors. The comparative analysis of the models demonstrated that the one-pole Cole-Cole model achieved a better fit compared to the one-pole Debye model across all bone types. Additionally, the two-pole Cole-Cole model exhibited a better fit than the two-pole Debye model. The proposed models can be used in numerical simulations to investigate the interaction of EM fields with bone tissue and develop and optimise microwave bone health monitoring and treatment techniques.

## I. INTRODUCTION

The dielectric properties, namely, the relative permittivity ($\varepsilon_r$) and conductivity ($\sigma\ (S/m)$) of biological tissues determine the dissipation of electromagnetic (EM) energy in the human body [1], [2]. When EM waves interact with biological tissues, they are partially absorbed and partially reflected. The amount of absorption and reflection depends on the dielectric properties of the tissue [3]. Moreover, these properties play a vital role in a wide variety of biomedical applications, including microwave imaging (MWI), microwave ablation (MWA), dosimetry studies, and the calculation of the Specific Absorption Rate (SAR) in biological tissues [2], [4], [5], [6], [7]. Therefore, understanding the dielectric properties of biological tissues and accurately modelling them is critical for advancing our understanding of the interaction of EM waves with the human body [8], [9]. To this end, both Cole-Cole and Debye models provide a framework for understanding the complex dielectric behaviour exhibited by biological materials [10], [11], [12]. This includes the polarisation and relaxation dynamics of water molecules, proteins, lipids, and other constituents present in tissues [13]. Moreover, these models facilitate the interpretation of dielectric measurements and aid in the development of accurate tissue models for biomedical simulations [14], [15], [16], [17].

The need for accurate modelling of EM interactions with biological tissue is crucial for the research and development of electromagnetic medical imaging and sensing (EMIS) techniques. MWI is an emerging diagnostic technology being investigated for a range of medical applications, including breast cancer detection and brain stroke monitoring [8]. The finite-difference time-domain (FDTD) method is a widely used computational technique for investigating MWI [18], [14], [19]. The FDTD simulations have been used to study various aspects of MWI, including the optimisation of antenna design and the evaluation of image reconstruction algorithms [8], [19], [20], [21], [22], [23], [24]. Moreover, the FDTD simulations are used for solving the forward problem in MWI, which involves predicting the EM field distribution given a known distribution of dielectric properties of the biological tissues. Further, these simulations play a key role in inverse scattering analysis, where the goal is to reconstruct the dielectric properties of an object from measured EM fields [8], [11]. The iterative development of highly sensitive imaging algorithms based on machine learning techniques requires extensive datasets, which can be challenging to obtain through physical experiments due to their inherent limitations and resource-intensive nature. Consequently, researchers rely on simulations to generate the requisite data, thereby enabling accelerated advancements in both imaging sensitivity and the effective integration of machine learning techniques. While EM simulation tools can employ polynomial curve fitting to simplify the modelling of frequency-dependent dielectric behaviour in biological tissues based on available measurement data, parametric models like the Debye model are crucial for capturing the intricate dielectric relaxation mechanisms present in biological tissues. These models are particularly essential for applications like microwave imaging, where they enable the reconstruction of tissue properties using inversion algorithms based on model parameters, unlike polynomial fit coefficients, which lack any physical interpretation.

To accurately model the dielectric properties of the biological tissues, FDTD methods incorporate various relaxation dispersion models of varying computational complexity [11]. While the Cole-Cole model provides the most accurate representation of the relaxation behaviour of biological tissues, the simplest relaxation dispersion model often used in FDTD simulations is the one-pole Debye model, which requires the least computational overhead and is widely used in simulations of microwave interactions with biological tissues [4], [14]. Debye models are often incorporated in these simulations as they provide a precise representation of how EM fields interact with biological tissue, which is critical for diagnostic imaging, therapy planning, and biomedical device design [25].

Studies have reported the Cole-Cole and Debye models to various biological tissues using different optimisation algorithms. Lazebnik *et al.* [14] fitted one- and two-pole Debye models on healthy and malignant breast tissues. Salahuddin *et al.* [4] developed multi-pole Debye parameters for 54 types of human tissues. Shahzad *et al.* [26] fitted a two-pole Cole-Cole model to the bovine adrenal gland dielectric properties. Fujii [11] fitted a three-pole Debye model to 20 human biological tissues. Dima *et al.* [27] fitted the Debye and Cole-Cole models to the dielectric properties of blood plasma with two different glucose concentrations. These models have been used in several studies across a wide range of applications, allowing researchers to adopt standardised models, facilitating reproducibility and consistency [4], [7], [14]. By using the same model parameters, different research groups can develop and validate methods on a common basis, leading to more reliable and comparable results. This uniformity is crucial for advancing the field as it ensures that findings and algorithms are developed on a standardised dataset, reducing variability due to differences in modelling approaches. Publishing these models not only fills these gaps but also provides a foundation for future research, enabling the community to build upon existing knowledge and improve the accuracy and effectiveness of imaging and diagnostic methods. Thus, the publication of these models is essential for promoting standardised practices and advancing the collective understanding of tissue dielectric properties.

While the dielectric properties of diseased human bones have been reported in the literature [28], no existing study has reported dielectric models for diseased human trabecular bones that can be incorporated in FDTD methods for modelling the diseased human bone. This gap is significant because accurate dielectric models with minimal errors are essential for precise EM simulations in microwave bone imaging for bone disease monitoring.

Therefore, this paper aims to fill this gap by developing and validating Cole-Cole and Debye models for diseased human trabecular bones, providing a valuable resource for various medical applications such as diagnostic imaging and sensor-enabled bone implants [29], [30]. The parameters reported in this work allow for computational modelling, numerical simulations, and device optimisation without requiring each group to re-derive these parameters. This accelerates innovation and improves consistency in the field. Dielectric properties' reporting using the mathematical models, such as Debye and Cole-Cole models, has been recommended in the study by Farrugia *et al.* [31] to facilitate the reproducibility of data and incorporation into numerical simulations. Following the recommendation of data reporting by Farrugia *et al.* [31], the fitting technique used to obtain the model, the root mean squared error of the models, and the 95% confidence intervals of the fitted parameters are also reported.

To this end, this study has developed the one-pole and two-pole Cole-Cole and Debye models for capturing the frequency-dependent dielectric properties of cortical bone and normal and diseased human trabecular bones over a frequency range of 0.5 – 8.5 GHz. The dielectric properties for cortical bone and normal trabecular bone were acquired from the parametric model presented in the study of Gabriel *et al*. [32] which used the experimentally measured literature data and its own experimental data for body tissues, whereas the measured dielectric properties of diseased human trabecular bone (osteoporosis and osteoarthritis) were acquired from Amin *et al*. [28], each at 101 linearly spaced frequency points. This work introduces, for the first time in the literature, dielectric models specifically Cole–Cole and Debye parameters for diseased human trabecular bone. These models will serve as a foundation for future studies in microwave-based bone imaging and diagnostics.

The remainder of the paper is organised as follows: Section 2 describes the methodology used in this study to develop the dielectric models for cortical bone and normal and diseased human trabecular bones. Section 3 presents the results and discussion of the developed one-pole and two-pole Cole-Cole and Debye models for cortical bone and normal and diseased human trabecular bones, comparing their accuracy to the existing experimental data, and discusses the implications of the developed models in medical imaging for bone health monitoring. Finally, conclusions are drawn in Section 4.

## II. METHODOLOGY

### A. Dielectric Models

The dielectric properties of biological tissues are inherently frequency-dependent [33]. Among the available dielectric models, the Cole-Cole and Debye models are the most extensively utilised for characterising these properties [4], [34], [35]. Both models describe the relaxation phenomena occurring within the dielectric spectrum of biological tissues [32]. However, due to the complex structural and compositional nature of biological tissues, the relaxation regions are typically broadened by multiple overlapping processes. While each relaxation process is defined by a characteristic relaxation time constant (τ), the Cole-Cole model introduces an additional distribution parameter (α) that accounts for the broadening of relaxation regions [32]. As a result, the Cole-Cole model generally offers a more accurate representation of the dielectric spectrum of biological tissues than the Debye model. The Cole-Cole model is defined as follows [36],

$$\hat{\varepsilon} = \varepsilon_\infty + \sum_{p=1}^{n} \frac{\Delta\varepsilon_p}{1+(j\omega\tau_p)^{(1-\alpha_p)}} + \frac{\sigma_s}{j\omega\epsilon_o} \qquad (1)$$

where $\hat{\varepsilon}$ is the complex permittivity of the material, $\varepsilon_\infty$ is the permittivity at the highest frequency, $\Delta\varepsilon_p$ is the change in the permittivity at the $p^{th}$ dispersion, $\tau_p$ is the relaxation constant at the $p^{th}$ dispersion, $\sigma_s$ is the static ionic conductivity, $\omega$ is the angular frequency, $\epsilon_o$ is the permittivity of free space and $\alpha_p$ is the distribution parameter at the $p^{th}$ dispersion, where $0 \leq \alpha \leq 1$.

The Cole-Cole model is computationally expensive and cannot be easily expressed in the time domain [4]. However, the Debye model can be easily expressed in both time and frequency domains and is not computationally expensive. The Debye parameters are widely used in EM numerical simulations [4]. The Debye model is defined as [37],

$$\hat{\varepsilon} = \varepsilon_\infty + \sum_{p=1}^{n} \frac{\Delta\varepsilon_p}{1+j\omega\tau_p} + \frac{\sigma_s}{j\omega\epsilon_o} \qquad (2)$$

The main difference between the Debye and Cole-Cole equations is that the latter includes the exponent $(1-\alpha)$.

### B. Numerical fitting technique

The numerical fitting technique used for parameter optimisation in this study is the weighted least squares method (W-LSM). The one-pole and two-pole Cole-Cole and Debye model parameters were optimised by using a W-LSM developed in MATLAB (R2024b, The MathWorks, Inc., Natick, MA, US). The W-LSM technique, originally proposed by Fujii [11], introduces a complex weight factor $\gamma$ to allow certain points to be emphasised or de-emphasised during fitting, depending on their relative importance or reliability. In its general form, the error weighting is expressed as:

$$\dot{E}_i \approx \{\hat{\varepsilon}(\omega_i)\}^\gamma \qquad (3)$$

where $\hat{\varepsilon}(\omega_i)$ is the measured complex permittivity at the frequency $\omega_i$. In this study, the weight factor $\gamma$ was set to 1, assigning equal weight to all permittivity values across the frequency range to ensure uniform fitting accuracy. The non-linear least squares method was used with the Newton iterative method to minimise the total weighted squared error, expressed as follows [4]:

$$E^2 = \sum_{i=1}^{N_f} \left[ \frac{\{C_r(\omega_i) - M_r(\omega_i)\}^2}{\{E_r(\omega_i)\}^2} + \frac{\{C_i(\omega_i) - M_i(\omega_i)\}^2}{\{E_i(\omega_i)\}^2} \right] \qquad (4)$$

where $N_f$ denotes the number of frequency points, $C_r(\omega_i)$ and $C_i(\omega_i)$ are the real and imaginary parts of the calculated complex permittivity, respectively. $M_r(\omega_i)$ and $M_i(\omega_i)$ are the real and imaginary parts of the measured complex permittivity, respectively, and $E_r(\omega_i)$ and $E_i(\omega_i)$ are the real and imaginary parts of the allowable error $\dot{E}_i$, respectively.

Both the Cole-Cole and Debye models were fitted to the mean, upper, and lower standard errors of the previously reported dielectric measurements of Amin *et al.* [28], where dielectric properties were measured for patients diagnosed with osteoporosis (*N = 7*) and osteoarthritis (*N = 5*) over the microwave frequency range (0.5 – 8.5 GHz). The standard error in [28] reflects intra-patient variability; therefore, model fits to both the upper and lower bounds were included to capture this range of variation in dielectric properties. The authors reported that the dielectric properties of trabecular bone vary significantly within the patient due to the heterogeneous microarchitecture. This variation in relative permittivity and conductivity was found to be 25% and 1.4%, respectively [28]. Moreover, the dielectric properties of osteoarthritis patients are significantly higher than osteoporotic patients, with an increase of 41% and 45% for relative permittivity and conductivity, respectively [28]. In addition, the models were fitted to the dielectric properties of normal trabecular and cortical bone dielectric properties data acquired from Gabriel *et al.* [32]. Gabriel *et al.'s* [32] study is considered the most comprehensive study and is widely used for characterising the dielectric properties of tissue. While the dataset for the dielectric properties of diseased human trabecular bones may be limited in size, its significance lies in being the only available dataset of its kind in the literature. Moreover, obtaining human trabecular bone samples, especially from healthy individuals, for experimental studies presents significant challenges due to ethical and logistical considerations.

### C. Performance metrics

The accuracy of the fitted dielectric models was assessed in terms of Root Mean Squared Error (RMSE), Frequency-dependent error, and Mean Absolute Percentage Error (MAPE).

1. ***Root Mean Squared Error:*** The total error between the measured and calculated dielectric properties resulting from the Cole-Cole and Debye models over the measured frequency range (0.5 – 8.5 GHz) was quantified using the RMSE. The RMSE is a commonly used metric that measures the magnitude of difference between predicted values and actual values. It represents the square root of the average of squared differences between predicted and actual values and therefore indicates how well the model fits the data. The RMSE can be calculated as follows:

$$RMSE = \sqrt{\frac{1}{n} \times (\sum(\varepsilon_i - \hat{\varepsilon}_i)^2)} \quad (5)$$

where $n$ denotes the number of data points, $\varepsilon_i$ is the $i_{th}$ actual value, $\hat{\varepsilon}_i$ is the $i_{th}$ predicted value, and $\Sigma$ denotes the sum of the squared differences between predicted and actual values.

2. ***Frequency-dependent error:*** Further, the frequency-dependent error between the experimental and calculated dielectric properties over the frequency range of 0.5 – 8.5 GHz was also investigated. The frequency-dependent errors are defined as follows:

$$E_{\varepsilon_r} = \varepsilon_r(\omega)_{cal} - \varepsilon_r(\omega)_{exp} \quad (6)$$

$$E_\sigma = \sigma(\omega)_{cal} - \sigma(\omega)_{exp} \quad (7)$$

where $E_{\varepsilon_r}$ is the difference between the calculated relative permittivity $\varepsilon_r(\omega)_{cal}$ and the experimental relative permittivity $\varepsilon_r(\omega)_{exp}$ at the frequency $\omega$. Similarly, $E_\sigma$ is the difference between the calculated conductivity $\sigma(\omega)_{cal}$ and the experimental conductivity $\sigma(\omega)_{exp}$ at the frequency $\omega$.

3. ***Mean Absolute Percentage Error:*** The prediction accuracy of both models is compared in terms of MAPE. MAPE represents the average percentage by which the predictions are far off the actual values. The MAPE can be calculated as follows:

$$MAPE = \frac{1}{n} \sum_{i=1}^{n} \left|\frac{\varepsilon_i - \hat{\varepsilon}_i}{\varepsilon_i}\right| \times 100 \quad (8)$$

where $n$ denotes the number of data points, $\varepsilon_i$ is the ith actual value, $\hat{\varepsilon}_i$ is the ith predicted value, and $\Sigma$ denotes the sum of the absolute percentage errors between predicted and actual values.

## III. RESULTS AND DISCUSSION

This section presents the results of the one-pole and two-pole Cole-Cole and Debye models for the cortical bone and normal and diseased human trabecular bones. The experimental and calculated dielectric properties from the Cole-Cole and Debye models are compared, and the errors are quantified.

### *A. One-pole Cole-Cole model for cortical bone and normal and diseased human trabecular bones*

The parameters of the one-pole Cole-Cole model were fitted to the mean, lower standard error, and upper standard error of the experimental dielectric properties of human trabecular bones affected by osteoporosis and osteoarthritis, as well as cortical and normal trabecular bone. Fig. 1 (a) and (b) show the relative permittivity and conductivity of the one-pole Cole-Cole model from 0.5 – 8.5 GHz, respectively. The curves of the one-pole Cole-Cole model (solid lines) demonstrate a good agreement with the experimental data

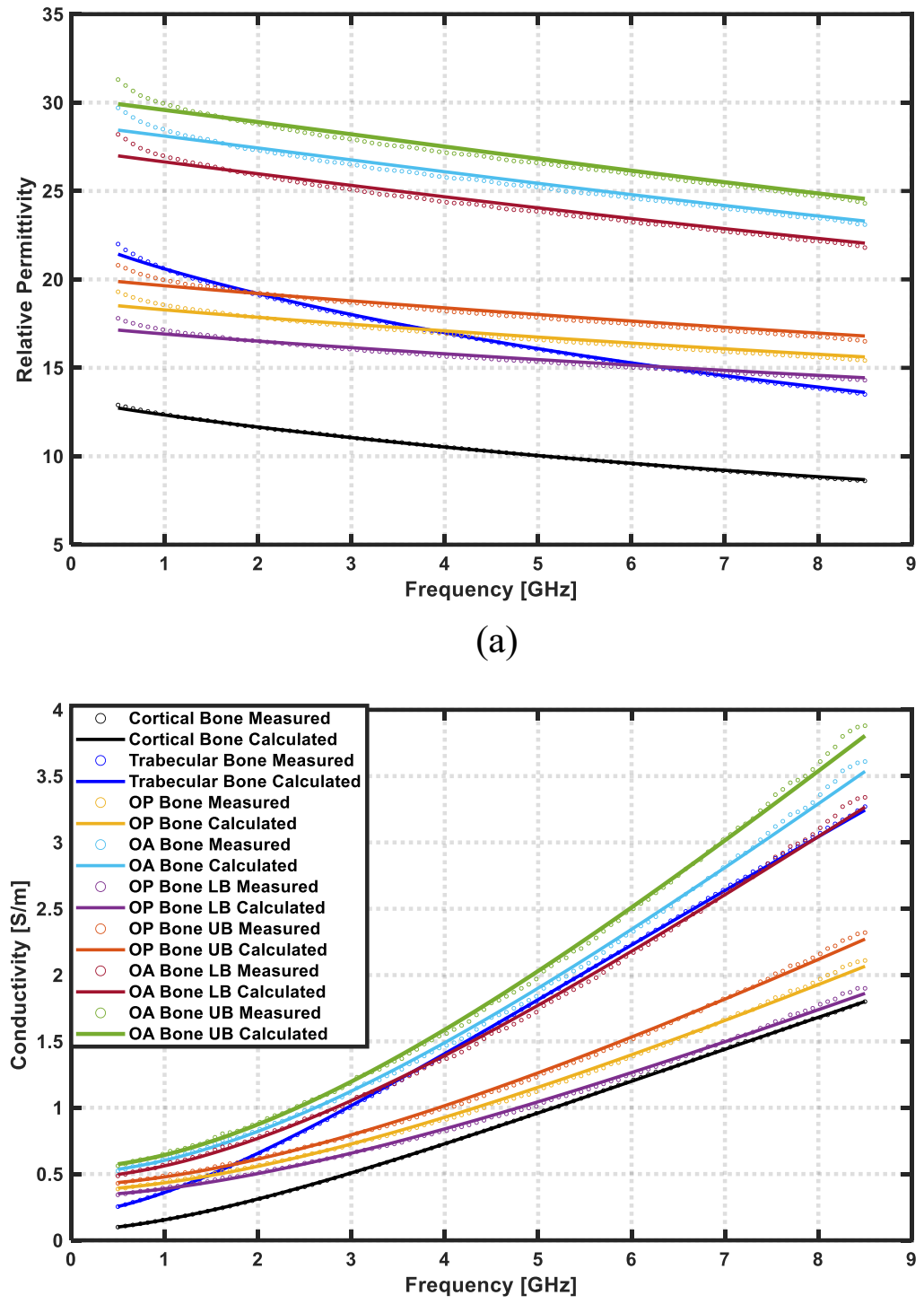

(a)

(b)

**FIGURE 1. One-pole Cole-Cole model parameters of all bones over 0.5 – 8.5 GHz frequency band: (a) Relative Permittivity (b) Conductivity. The experimental dielectric data of bones (scatter points) are compared with calculated data from the one-pole Cole-Cole model (solid lines). OP = Osteoporotic, OA = Osteoarthritis, LB = Lower Bound, UB = Upper Bound.**

(scatter points) over the entire frequency band. The distribution parameter α in the Cole-Cole model, which accounts for the broadening of the relaxation region, provided a good fit in almost the entire frequency band. However, a slight difference can be observed between the experimental and calculated relative permittivity data at lower frequencies (0.5 – 1.5 GHz) for the diseased trabecular bones. The parameters of the one-pole Cole-Cole model fitted to the mean, lower standard error, and upper standard error dielectric properties of osteoporotic and osteoarthritis trabecular bones, as well as the cortical bone and normal trabecular bone, are presented in Table 1.

**TABLE 1. One-pole Cole-Cole model parameters of all investigated bones with 95% confidence intervlas *Δ*.**

| Bone Type | $\varepsilon_\infty \pm \Delta$ | $\sigma_s \pm \Delta$ (S/m) | $\Delta\varepsilon \pm \Delta$ | $\tau$ $\pm\Delta$ (ps) | $\alpha \pm \Delta$ |
|---|---|---|---|---|---|
| Cortical Bone | 1.08 ± 0.34 | 0.07 ± 0.001 | 12.20 ± 0.43 | 11.28 ± 0.75 | 0.29 ± 0.01 |
| Trabecular Bone | 1.00 ± 0.62 | 0.20 ± 0.004 | 21.64 ± 1.11 | 13.17 ± 1.09 | 0.31 ± 0.02 |
| OP Bone Mean | 5.00 ± 3.56 | 0.38 ± 0.007 | 13.82 ± 3.27 | 5.97 ± 2.35 | 0.26 ± 0.07 |
| OP Bone LB | 5.00 ± 2.72 | 0.34 ± 0.005 | 12.44 ± 2.56 | 6.15 ± 2.16 | 0.27 ± 0.06 |
| OP Bone UB | 5.00 ± 4.89 | 0.42 ± 0.009 | 15.20 ± 4.34 | 5.84 ± 2.67 | 0.25 ± 0.09 |
| OA Bone Mean | 8.00 ± 3.28 | 0.51 ± 0.009 | 20.83 ± 3.21 | 8.07 ± 1.98 | 0.20 ± 0.05 |
| OA Bone LB | 8.00 ± 2.95 | 0.47 ± 0.009 | 19.39 ± 2.92 | 8.21 ± 2.00 | 0.21 ± 0.05 |

| OA Bone UB | 8.00 ± 3.64 | 0.55 ± 0.010 | 22.29 ± 3.54 | 7.96 ± 1.99 | 0.19 ± 0.05 |
|---|---|---|---|---|---|

OP = Osteoporotic, OA = Osteoarthritis, LB = Lower Bound, UB = Upper Bound.

The RMSE between the experimental and calculated dielectric properties data from the one-pole Cole-Cole model is presented in Table 2. Fig. 2 (a) and (b) show the frequency-dependent error in the relative permittivity and conductivity of the experimental and calculated dielectric properties data from 0.5 – 8.5 GHz for the one-pole Cole-Cole model, respectively. In Fig. 2 (a), due to a slight difference between the experimental and calculated relative permittivity of the diseased trabecular bones at lower frequencies (0.5 – 1.5 GHz), there is an error reaching up to 1.3. In the remaining frequency band, the error is centred around 0. Moreover, in Fig. 2 (b), the frequency-dependent error between the experimental and calculated conductivity is centred around 0 from 0.5 – 8.5 GHz for all the bone types.

### *B. Two-pole Cole-Cole model for cortical bone and normal and diseased human trabecular bones*

Although the one-pole Cole-Cole model fitted very well to the experimental dielectric properties of all the bone types in the frequency range 0.5 – 8.5 GHz, the two-pole Cole-Cole model was fitted to investigate its behaviour in comparison with the one-pole Cole-Cole model for the biological tissues under study. Fig. 3 (a) and (b) show the relative permittivity and conductivity of the two-pole Cole-Cole model from 0.5 – 8.5 GHz, respectively. The curves obtained from the two-pole Cole-Cole model (solid lines) demonstrate an excellent agreement with the experimental data (scatter points) over the entire 0.5 – 8.5 GHz frequency band. Overall, the behaviour of the two-pole Cole-Cole model is significantly improved from the one-pole Cole-Cole model, especially at lower frequencies (0.5 – 1.5 GHz) for relative permittivity. The parameters of the two-pole Cole-Cole model fitted to the mean, lower standard error, and upper standard error dielectric properties of osteoporotic and osteoarthritis trabecular bones, as well as the cortical bone and normal trabecular bone, are presented in Table 3.

The RMSE between the experimental and calculated dielectric properties data from the two-pole Cole-Cole model is presented in Table 4. Fig. 4 (a) and (b) show the frequency-dependent error in the relative permittivity and conductivity of the experimental and calculated dielectric properties data from 0.5 – 8.5 GHz for the two-pole Cole-Cole model, respectively.

**TABLE 2.** **Root Mean Squared Error values of the one-pole Cole-Cole model.**

| Bone Type | Root Mean Squared Error (RMSE) | |
|---|---|---|
| | $\varepsilon_r$ | $\sigma$ (S/m) |
| Cortical Bone | 0.0424 | 0.0027 |
| Trabecular Bone | 0.0985 | 0.0113 |
| OP Bone Mean | 0.1929 | 0.0194 |
| OP Bone LB | 0.1597 | 0.0181 |
| OP Bone UB | 0.2317 | 0.0216 |
| OA Bone Mean | 0.2915 | 0.0346 |
| OA Bone LB | 0.2760 | 0.0332 |
| OA Bone UB | 0.3079 | 0.0361 |

OP = Osteoporotic, OA = Osteoarthritis, LB = Lower Bound, UB = Upper Bound.

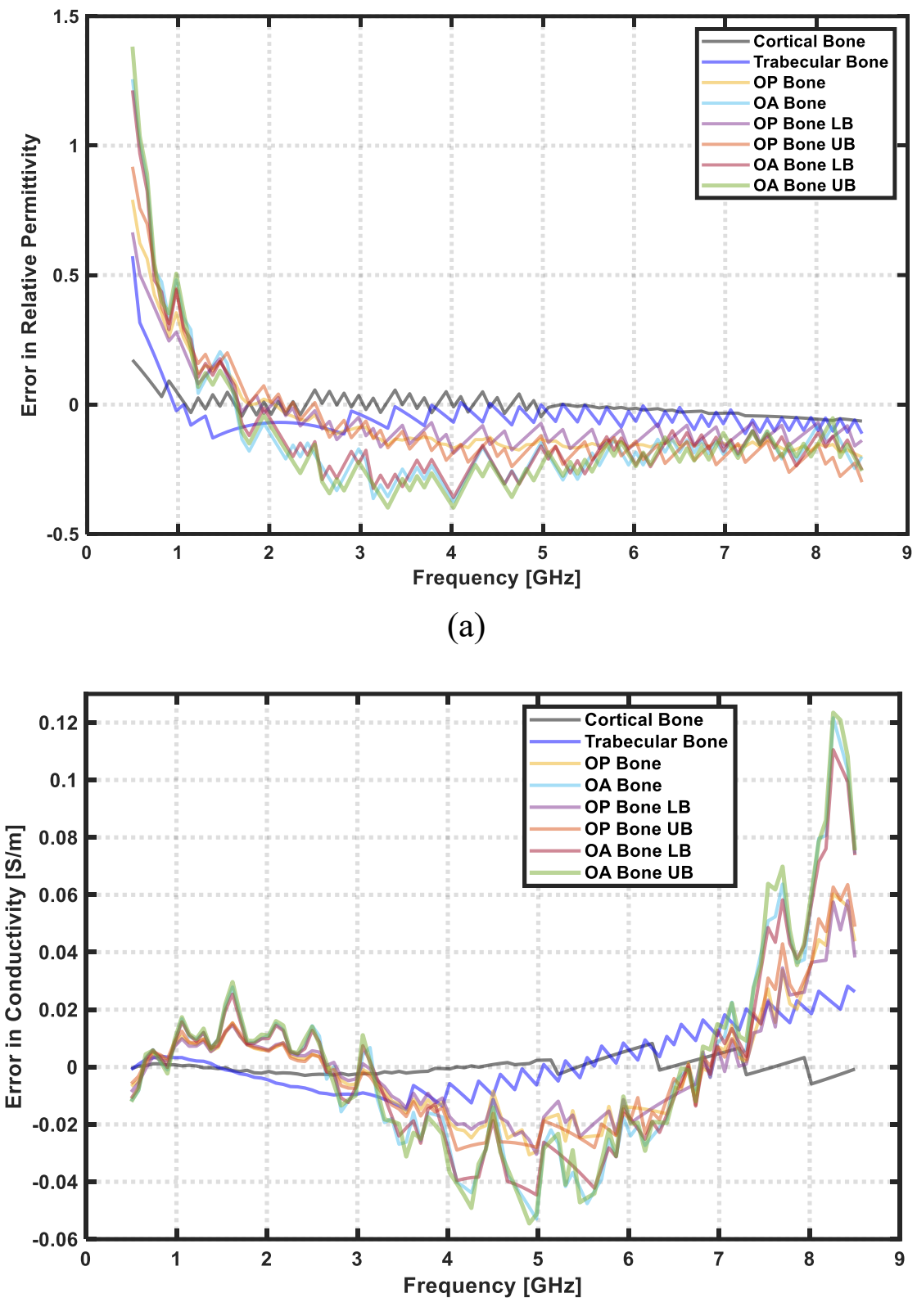

**FIGURE 2.** **Frequency-dependent error in one-pole Cole-Cole model parameters of all bones over 0.5 – 8.5 GHz frequency band: (a) Relative Permittivity (b) Conductivity. OP = Osteoporotic, OA = Osteoarthritis, LB = Lower Bound, UB = Upper Bound.**

**TABLE 3.** **Two-pole Cole-Cole model parameters of all investigated bones with 95% confidence intervals *Δ*.**

| B.T | $\varepsilon_\infty \pm \Delta$ | $\sigma_s \pm \Delta$ $(S/m)$ | $\Delta\varepsilon_1 \pm \Delta$ | $\tau_1 \pm \Delta$ (ps) | $\alpha_1 \pm \Delta$ | $\Delta\varepsilon_2 \pm \Delta$ | $\tau_2 \pm \Delta$ (ps) | $\alpha_2 \pm \Delta$ |
|---|---|---|---|---|---|---|---|---|
| P1 | 2.52 ± 0.12 | 0.06 ± 0.001 | 2.50 ± 0.00 | 804.33 ± 0.015 | 0.28 ± 0 | 9.97 ± 0.147 | 13.29 ± 0.28 | 0.20 ± 0.006 |
| P2 | 3.48 ± 0.24 | 0.18 ± 0.002 | 5.00 ± 0.00 | 300.00 ± 0.003 | 0.30 ± 0 | 16.15 ± 0.283 | 13.80 ± 0.31 | 0.17 ± 0.008 |
| P3 | 8.87 ± 0.39 | 0.32 ± 0.011 | 7.54 ± 0.56 | 616.91 ± 0.031 | 0.34 ± 0.09 | 7.84 ± 0.000 | 9.57 ± 0.68 | 0 |
| P4 | 7.99 ± 0.85 | 0.30 ± 0.003 | 4.05 ± 0.000 | 225.86 ± 0.04 | 0.25 ± 0 | 7.47 ± 0.863 | 8.84 ± 1.23 | 0 |
| P5 | 9.13 ± 1.26 | 0.35 ± 0.005 | 7.98 ± 0.002 | 580.86 ± 0.36 | 0.29 ± 0 | 8.97 ± 1.350 | 9.26 ± 1.59 | 0 |
| P6 | 10.34 ± 1.32 | 0.43 ± 0.003 | 13.44 ± 0.044 | 710.55 ± 5.29 | 0.44 ± 0 | 14.44 ± 1.380 | 8.92 ± 1.04 | 0 |
| P7 | 10.37 ± 1.17 | 0.39 ± 0.003 | 12.47 ± 0.043 | 616.38 ± 4.95 | 0.45 ± 0 | 12.93 ± 1.213 | 9.12 ± 1.04 | 0 |
| P8 | 10.57 ± 1.41 | 0.47 ± 0.003 | 11.52 ± 0.025 | 507.96 ± 2.37 | 0.40 ± 0 | 15.87 ± 1.469 | 8.90 ± 1.01 | 0 |

P1 = Cortical Bone, P2 = Trabecular Bone, P3 = Osteoporotic (OP) Bone Mean, P4 = OP Bone Lower Bound (LB), P5 = OP Bone Upper Bound (UB), P6 = Osteoarthritis (OA) Bone Mean, P7 = OA Bone LB, P8 = OA Bone UB.

**TABLE 4.** **Root Mean Squared Error values of the two-pole Cole-Cole model.**

| Bone Type | Root Mean Squared Error (RMSE) | |
|---|---|---|
| | $\varepsilon_r$ | $\sigma$ (S/m) |
| Cortical Bone | 0.0219 | 0.0019 |
| Trabecular Bone | 0.0387 | 0.0030 |
| OP Bone Mean | 0.0837 | 0.0092 |
| OP Bone LB | 0.0872 | 0.0063 |
| OP Bone UB | 0.1225 | 0.0086 |
| OA Bone Mean | 0.1233 | 0.0124 |
| OA Bone LB | 0.1208 | 0.0112 |
| OA Bone UB | 0.1237 | 0.0127 |

OP = Osteoporotic, OA = Osteoarthritis, LB = Lower Bound, UB = Upper Bound.

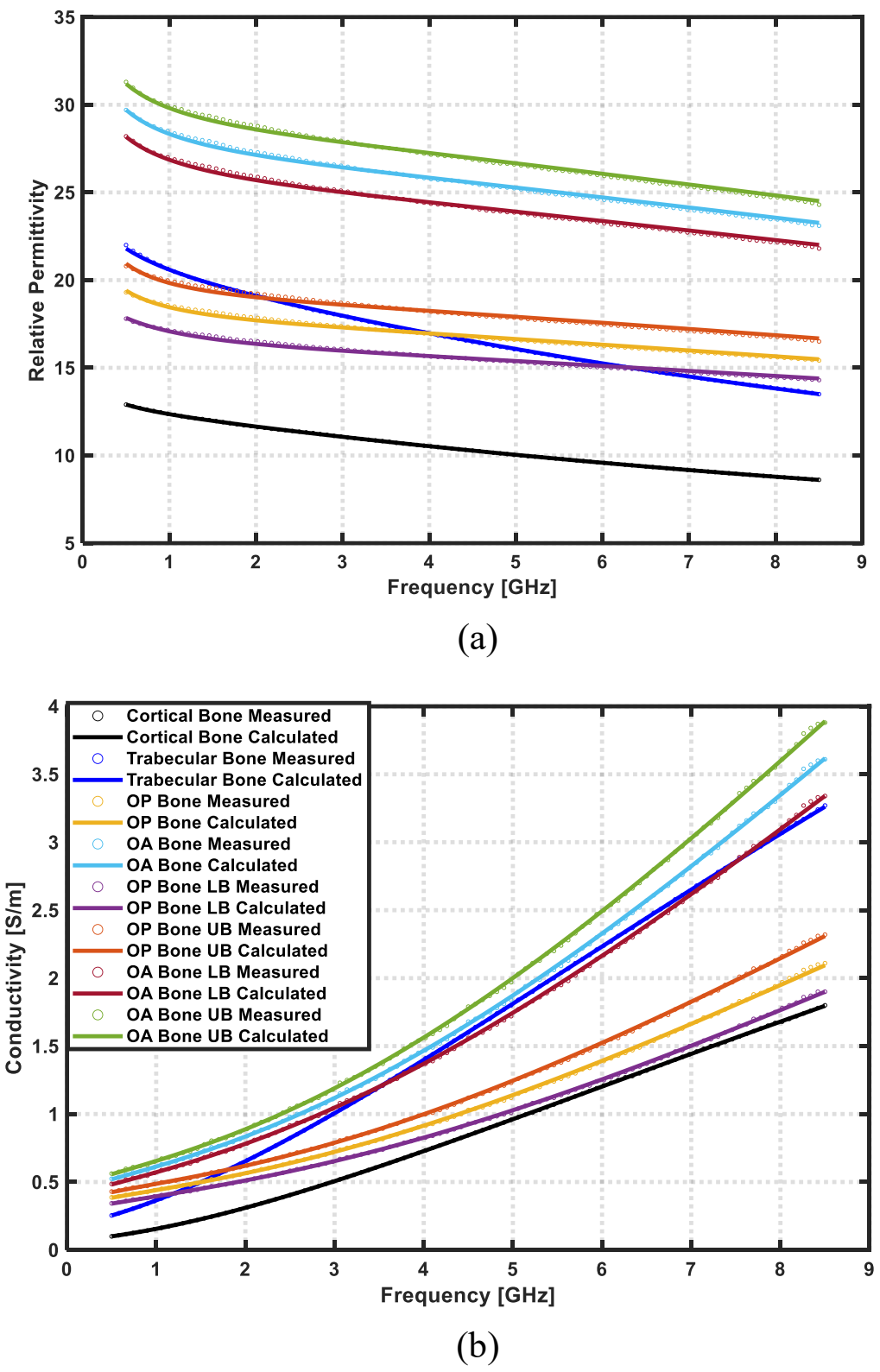

**FIGURE 3. Two-pole Cole-Cole model parameters of all bones over 0.5 – 8.5 GHz frequency band: (a) Relative Permittivity (b) Conductivity. The experimental dielectric data of bones (scatter points) are compared with calculated data from the one-pole Cole-Cole model (solid lines). OP = Osteoporotic, OA = Osteoarthritis, LB = Lower Bound, UB = Upper Bound.**

### *C. One-pole Debye model for cortical bone and normal and diseased human trabecular bones*

The fitting of the one-pole Debye model to the cortical bone and normal and diseased human trabecular bones was also incorporated into the study. Fig. 5 (a) and (b) show the relative permittivity and conductivity of the one-pole Debye model from 0.5 – 8.5 GHz, respectively. The curves obtained from the one-pole Debye model (solid lines) do not fit very well with the experimental data (scatter points). A notable difference can be observed between the experimental and calculated relative permittivity data at lower frequencies (0.5– 1.5 GHz). Additionally, at higher frequencies (2 – 6 GHz), the calculated values of relative permittivity from the one-pole Debye model are higher than the experimental values. One potential reason for the mismatch at low frequencies < 1.5 GHz and an overestimation of dielectric properties at higher frequencies (2 – 6 GHz) is that the one-pole Debye model, which assumes a single relaxation time, may not fully capture the intricate behaviour of more complex biological tissue such as bone at those frequencies. Additionally, biological tissues often exhibit more complex behaviour at lower frequencies (< 1.5 GHz), such as frequency-dependent conductivity or multi-relaxation phenomena, which the one-pole Debye model may not adequately represent [38], [39]. The one-pole Debye model parameters for the cortical bone and normal and diseased human trabecular bone are presented in Table 5.

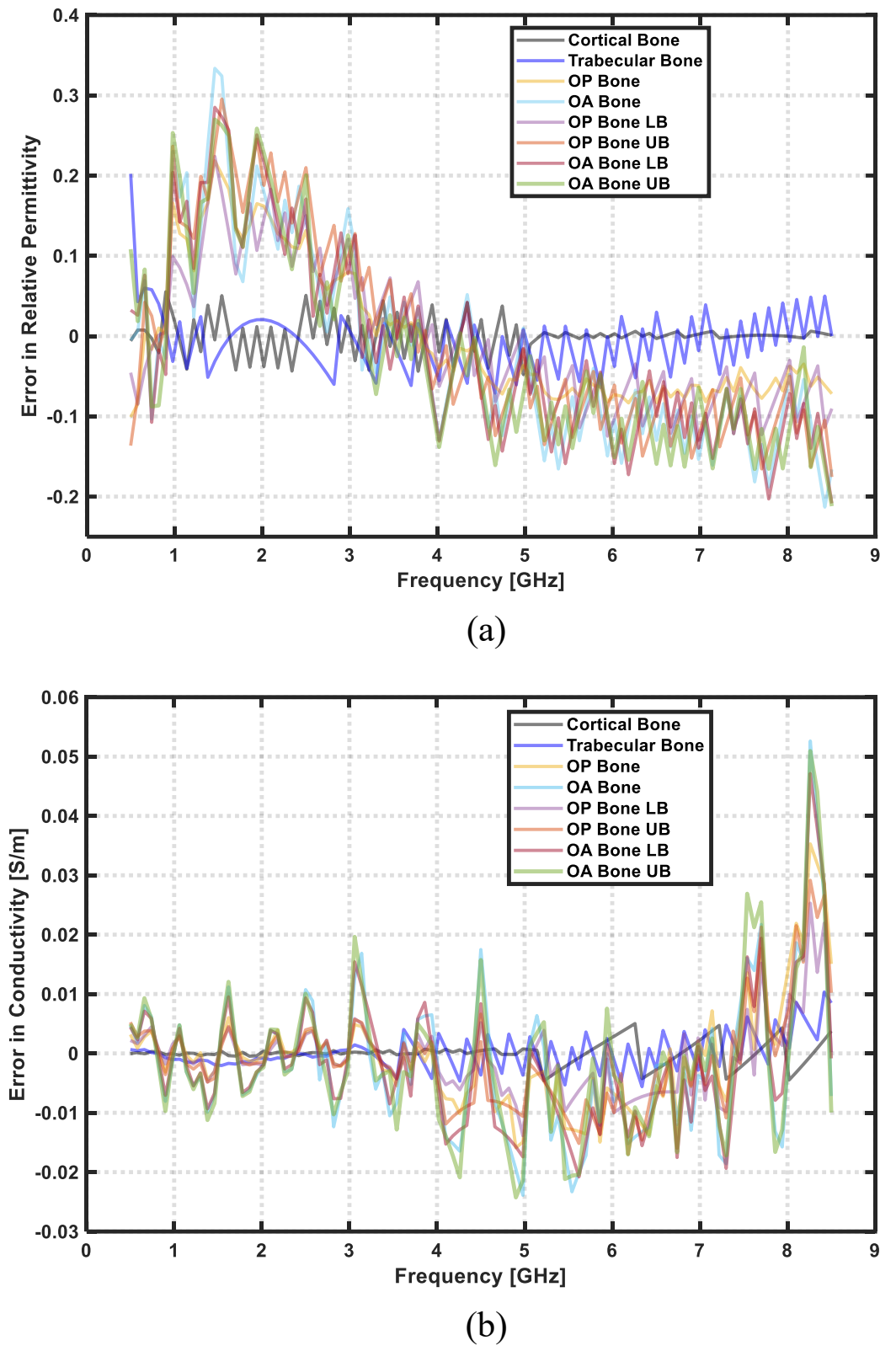

**FIGURE 4. Frequency-dependent error in two-pole Cole-Cole model parameters of all bones over 0.5 – 8.5 GHz frequency band: (a) Relative Permittivity (b) Conductivity. OP = Osteoporotic, OA = Osteoarthritis, LB = Lower Bound, UB = Upper Bound.**

**TABLE 5. One-pole Debye model parameters of all investigated bones with 95% confidence intervals *Δ*.**

| Bone Type | $\varepsilon_\infty \pm \Delta$ | $\sigma_s \pm \Delta$ (S/m) | $\Delta\varepsilon \pm \Delta$ | $\tau \pm \Delta$ (ps) |
|---|---|---|---|---|
| Cortical Bone | 5.52 ± 0.22 | 0.10 ± 0.002 | 6.93 ± 0.10 | 22.53 ± 0.61 |
| Trabecular Bone | 8.30 ± 0.39 | 0.25 ± 0.006 | 12.35 ± 0.21 | 23.57 ± 0.69 |
| OP Bone Mean | 11.20 ± 0.22 | 0.40 ± 0.006 | 7.04 ± 0.14 | 15.46 ± 0.83 |
| OP Bone LB | 10.55 ± 0.22 | 0.36 ± 0.005 | 6.31 ± 0.12 | 15.92 ± 0.78 |
| OP Bone UB | 11.83 ± 0.23 | 0.44 ± 0.007 | 7.77 ± 0.17 | 15.05 ± 0.89 |
| OA Bone Mean | 15.37 ± 0.41 | 0.54 ± 0.009 | 12.68 ± 0.24 | 15.52 ± 0.68 |
| OA Bone LB | 14.97 ± 0.38 | 0.50 ± 0.008 | 11.62 ± 0.21 | 16.05 ± 0.70 |
| OA Bone UB | 15.77 ± 0.43 | 0.58 ± 0.009 | 13.76 ± 0.27 | 15.08 ± 0.66 |

OP = Osteoporotic, OA = Osteoarthritis, LB = Lower Bound, UB = Upper Bound.

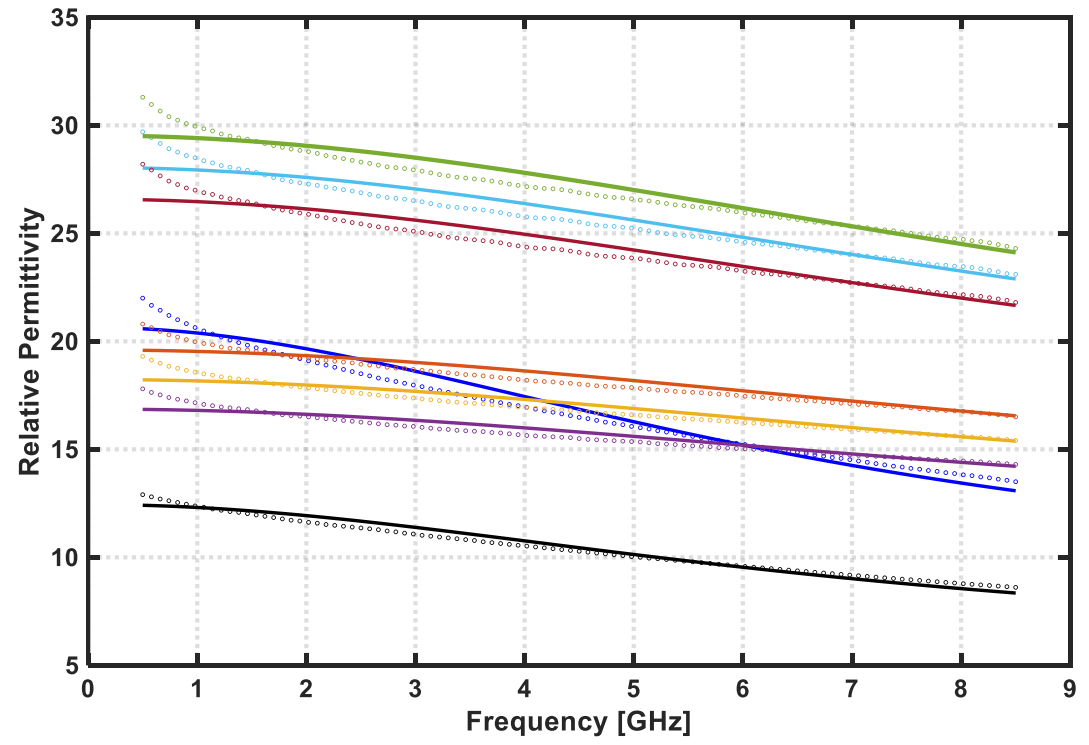

(a)

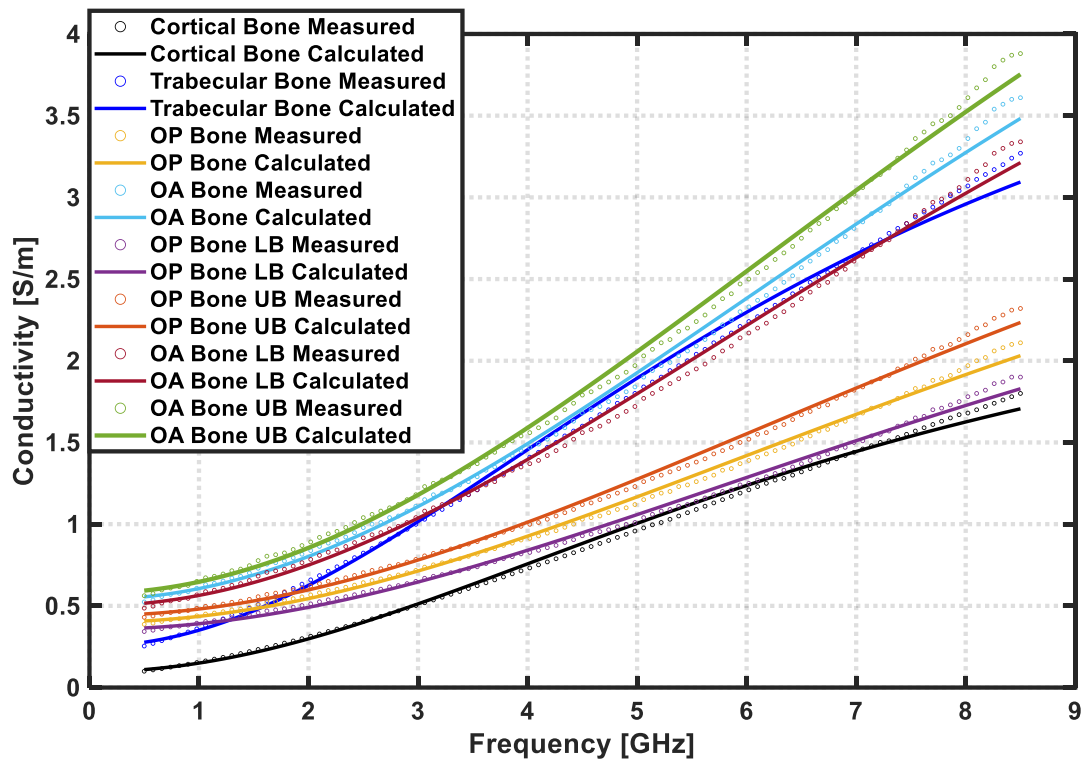

(b)

**FIGURE 5. One-pole Debye model parameters of all bones over 0.5 – 8.5 GHz frequency band: (a) Relative Permittivity (b) Conductivity. The experimental dielectric data of bones (scatter points) are compared with calculated data from the two-pole Debye model (solid lines). OP = Osteoporotic, OA = Osteoarthritis, LB = Lower Bound, UB = Upper Bound.**

The RMSE between experimental and calculated dielectric properties data from the one-pole Debye model is presented in Table 6. Fig. 6 (a) and (b) show the frequency-dependent error in the relative permittivity and conductivity of the experimental and calculated dielectric properties data of all investigated bones from 0.5 – 8.5 GHz for the one-pole Debye model, respectively. It can be observed from Fig. 6 (a) that for the one-pole Debye model, the error between the experimental and calculated relative permittivity is significantly larger for low frequencies (0.5 – 1 GHz), i.e., reaching up to 1.8 and for the higher frequencies (2 – 6 GHz), it is up to 0.7. Moreover, in Fig. 6 (b), the frequency-dependent error between the experimental and calculated conductivity is centred around 0 from 0.5 – 8.5 GHz for all the bone types. These results indicate that the one-pole Debye model may not accurately represent the relative permittivity of the investigated bones over the investigated frequency band.

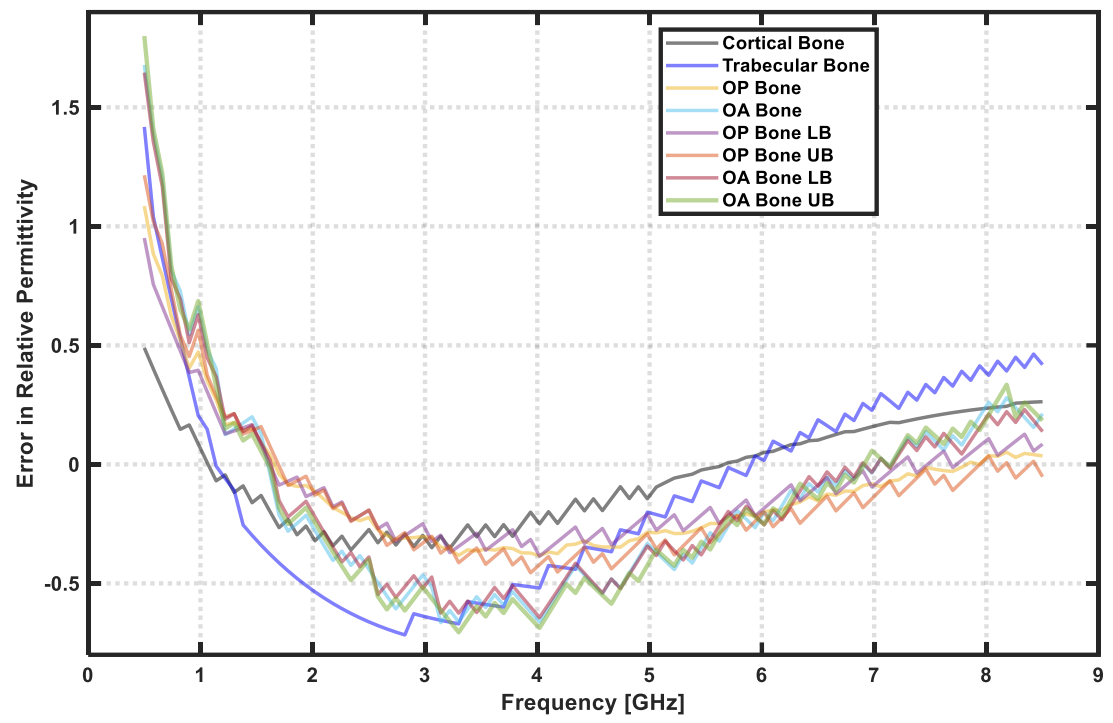

(a)

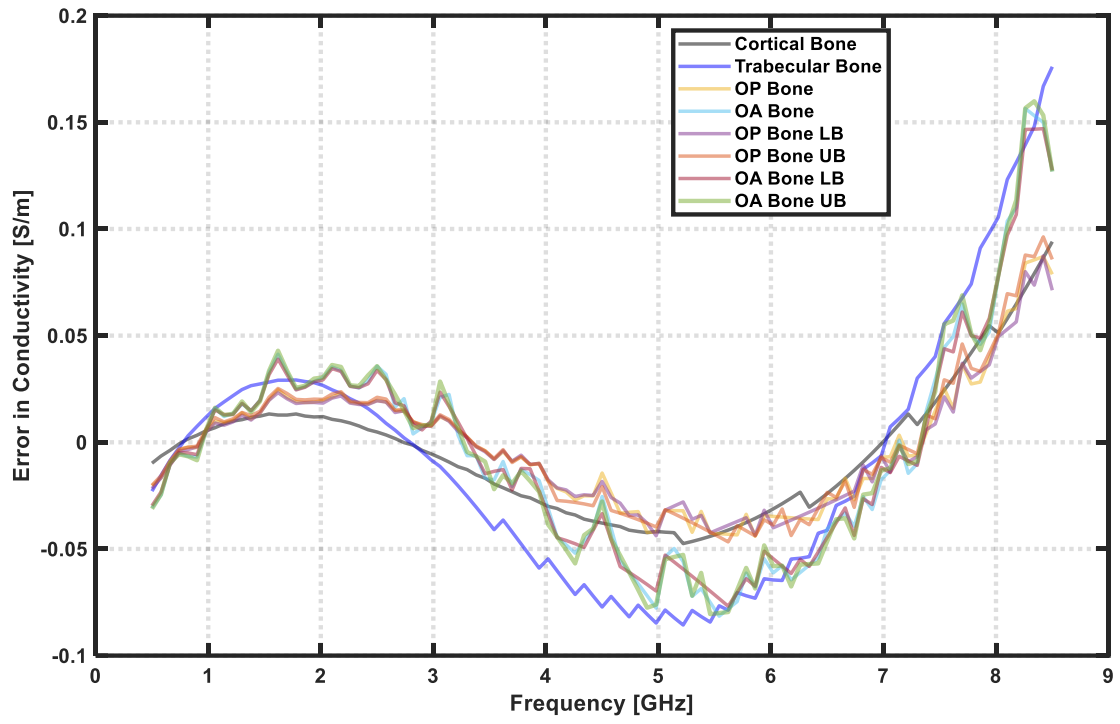

(b)

**FIGURE 6. Frequency-dependent error in one-pole Debye model parameters of all bones over 0.5 – 8.5 GHz frequency band: (a) Relative Permittivity (b) Conductivity. OP = Osteoporotic, OA = Osteoarthritis, LB = Lower Bound, UB = Upper Bound.**

**TABLE 6. Root Mean Squared Error values of the one-pole Debye model.**

| Bone Type | Root Mean Squared Error (RMSE) | |
|---|---|---|
| | $\varepsilon_r$ | $\sigma$ (S/m) |
| Cortical Bone | 0.2184 | 0.0316 |
| Trabecular Bone | 0.4502 | 0.0608 |
| OP Bone Mean | 0.2973 | 0.0301 |
| OP Bone LB | 0.2638 | 0.0289 |
| OP Bone UB | 0.3353 | 0.0316 |
| OA Bone Mean | 0.4572 | 0.0520 |
| OA Bone LB | 0.4402 | 0.0500 |
| OA Bone UB | 0.4749 | 0.0529 |

OP = Osteoporotic, OA = Osteoarthritis, LB = Lower Bound, UB = Upper Bound.

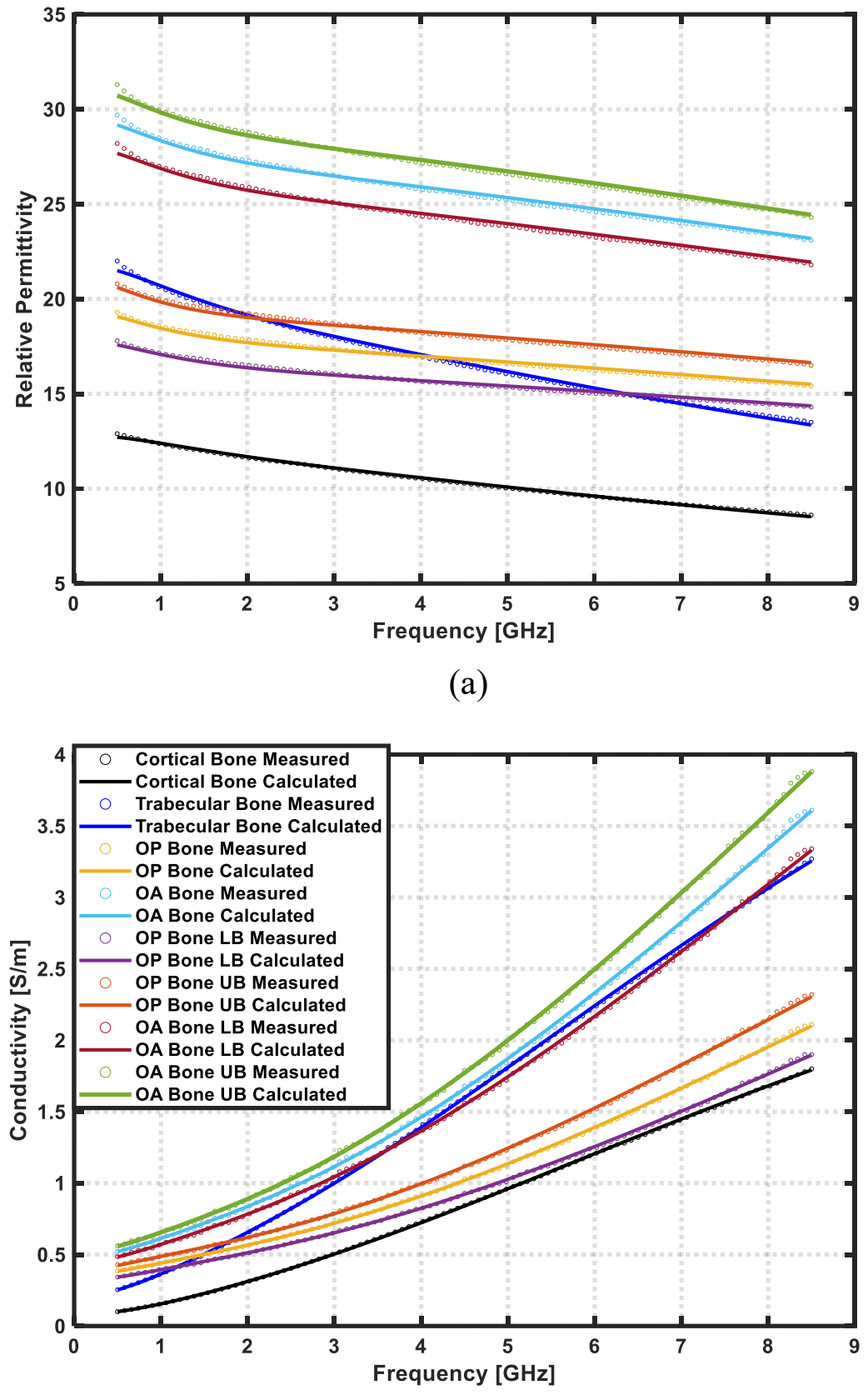

(b)

**FIGURE 7. Two-pole Debye model parameters of all bones over 0.5 – 8.5 GHz frequency band: (a) Relative Permittivity (b) Conductivity. The experimental dielectric data of bones (scatter points) are compared with calculated data from the two-pole Debye model (solid lines). OP = Osteoporotic, OA = Osteoarthritis, LB = Lower Bound, UB = Upper Bound.**

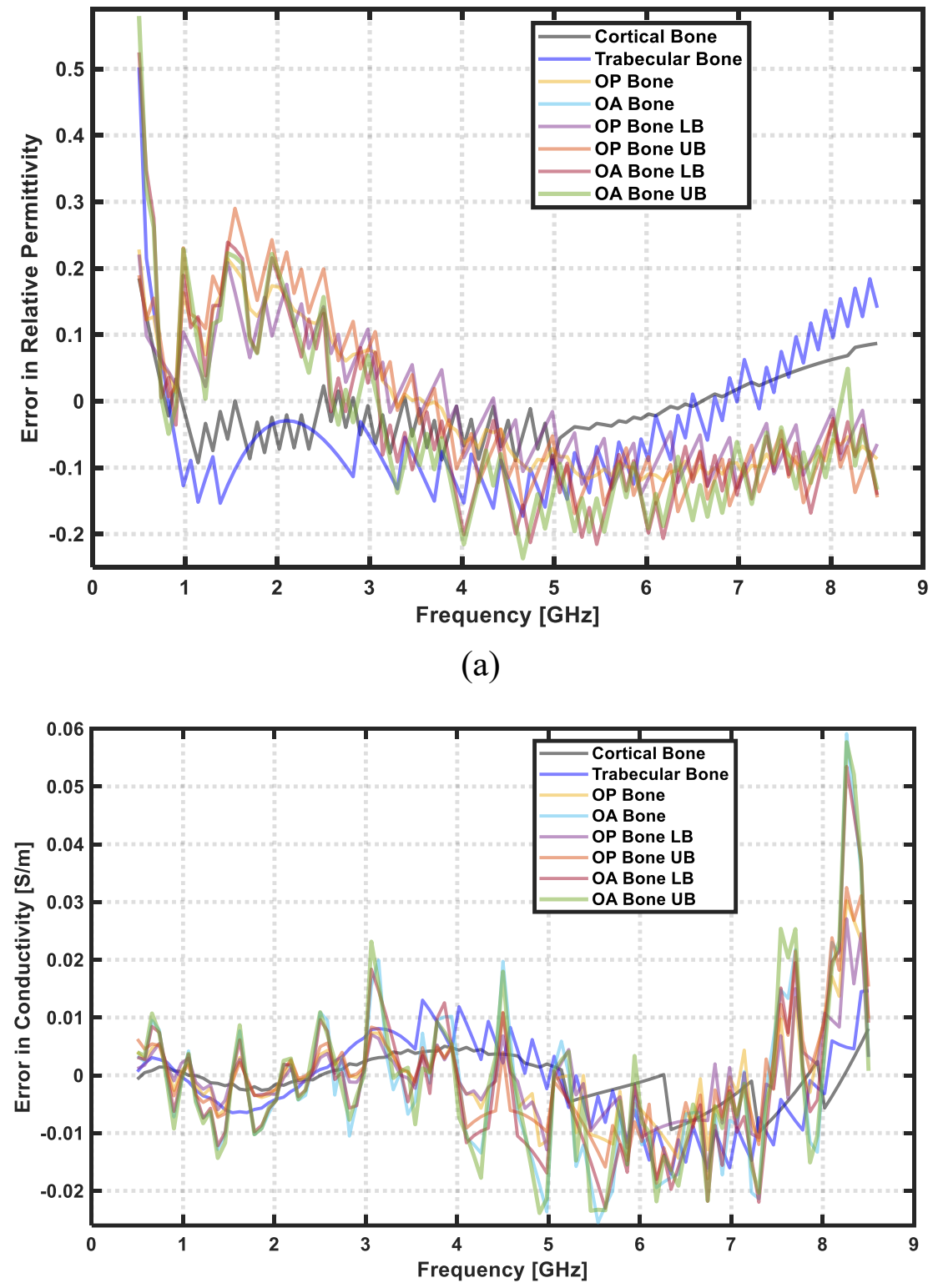

(b)

**FIGURE 8. Frequency-dependent error in two-pole Debye model parameters of all bones over 0.5 – 8.5 GHz frequency band: (a) Relative Permittivity (b) Conductivity. OP = Osteoporotic, OA = Osteoarthritis, LB = Lower Bound, UB = Upper Bound.**

### *D. Two-pole Debye model for cortical bone and normal and diseased human trabecular bones*

To overcome the limitations of the single-pole Debye model, this study investigated the two-pole Debye model that considers two relaxation times. By including additional relaxation times, the two-pole Debye model can provide a more accurate description of the dielectric properties of biological tissues, particularly at lower frequencies where the behaviour of the tissue may be more complex due to underlying confounders of polarisation [4], [26]. Like the one-pole Debye model, the parameters of the two-pole Debye model were fitted to the mean, lower standard error, and upper standard error of the experimental dielectric properties of human trabecular bones affected by osteoporosis and osteoarthritis, as well as cortical bone and normal trabecular bone dielectric properties. Fig. 7 (a) and (b) show the relative permittivity and conductivity of the two-pole Debye model from 0.5 – 8.5 GHz, respectively. The curves obtained from the two-pole Debye model (solid lines) demonstrate excellent agreement with the experimental data (scatter points) over the entire 0.5 – 8.5 GHz frequency band. This strong agreement indicates that the two-pole Debye model can accurately represent both the relative permittivity and conductivity of the investigated bones and is an improvement over the one-pole Debye model, which showed higher errors over the investigated frequency band. The parameters of the two-pole Debye model fitted to the mean, lower standard error, and upper standard error dielectric properties of osteoporotic and osteoarthritis trabecular

**TABLE 7. Two-pole Debye model parameters of all investigated bones with 95% confidence intervals *Δ*.**

| B.T | $\varepsilon_\infty \pm \Delta$ | $\sigma_s \pm \Delta$ (S/m) | $\Delta\varepsilon_1 \pm \Delta$ | $\tau_1 \pm \Delta$ (ps) | $\Delta\varepsilon_2 \pm \Delta$ | $\tau_2 \pm \Delta$ (ps) |
|---|---|---|---|---|---|---|
| P1 | 4.50 ± 0.10 | 0.08 ± 0.001 | 6.68 ± 0.05 | 15.58 ± 0.43 | 1.67 ± 0.06 | 88.14 ± 4.12 |
| P2 | 6.52 ± 0.24 | 0.21 ± 0.005 | 11.75 ± 0.15 | 16.19 ± 0.53 | 3.60 ± 0.16 | 102.30 ± 6.45 |
| P3 | 9.22 ± 0.29 | 0.35 ± 0.014 | 2.09 ± 0.43 | 145.46 ± 26.31 | 8.13 ± 0.23 | 10.27 ± 0.57 |
| P4 | 8.56 ± 0.26 | 0.32 ± 0.006 | 1.89 ± 0.18 | 128.68 ± 13.30 | 7.40 ± 0.21 | 9.99 ± 0.54 |
| P5 | 9.82 ± 0.36 | 0.39 ± 0.018 | 8.90 ± 0.29 | 10.43 ± 0.65 | 2.41 ± 0.62 | 165.26 ± 35.78 |
| P6 | 12.06 ± 0.46 | 0.48 ± 0.010 | 14.53 ± 0.38 | 10.47 ± 0.52 | 3.03 ± 0.30 | 129.74 ± 14.70 |
| P7 | 11.98 ± 0.42 | 0.44 ± 0.010 | 2.95 ± 0.27 | 128.37 ± 14.30 | 13.17 ± 0.33 | 10.77 ± 0.56 |
| P8 | 12.27 ± 0.50 | 0.51 ± 0.012 | 15.81 ± 0.41 | 10.36 ± 0.51 | 3.13 ± 0.34 | 134.79 ± 16.19 |

P1 = Cortical Bone, P2 = Trabecular Bone, P3 = Osteoporotic (OP) Bone Mean, P4 = OP Bone Lower Bound (LB), P5 = OP Bone Upper Bound (UB), P6 = Osteoarthritis (OA) Bone Mean, P7 = OA Bone LB, P8 = OA Bone UB.

**TABLE 8. Root Mean Squared Error values of the two-pole Debye model.**

| Bone Type | Root Mean Squared Error (RMSE) | |
|---|---|---|
| | $\varepsilon_r$ | $\sigma$ (S/m) |
| Cortical Bone | 0.0529 | 0.0038 |
| Trabecular Bone | 0.1105 | 0.0077 |
| OP Bone Mean | 0.1025 | 0.0079 |
| OP Bone LB | 0.0889 | 0.0067 |
| OP Bone UB | 0.1269 | 0.0094 |
| OA Bone Mean | 0.1414 | 0.0140 |
| OA Bone LB | 0.1396 | 0.0130 |
| OA Bone UB | 0.1449 | 0.0143 |

OP = Osteoporotic, OA = Osteoarthritis, LB = Lower Bound, UB = Upper Bound.

bones, as well as the cortical bone and normal trabecular bone, are presented in Table 7.

The RMSE between the experimental and calculated dielectric properties data from the two-pole Debye model is presented in Table 8. It can be observed from Table 8 that the RMSE for both relative permittivity and conductivity for the two-pole Debye model is lower when compared to the RMSE for relative permittivity and conductivity of the one-pole Debye model. Fig. 8 (a) and (b) show the frequency-dependent error in the relative permittivity and conductivity of the experimental and calculated dielectric properties data from 0.5 – 8.5 GHz for the two-pole Debye models, respectively. It can be observed from Fig. 8 (a) and (b) that for the two-pole Debye model, the error between the experimental and calculated relative permittivity is negligible for the entire band, due to which the frequency-dependent error is centred around 0 for all the investigated bones.

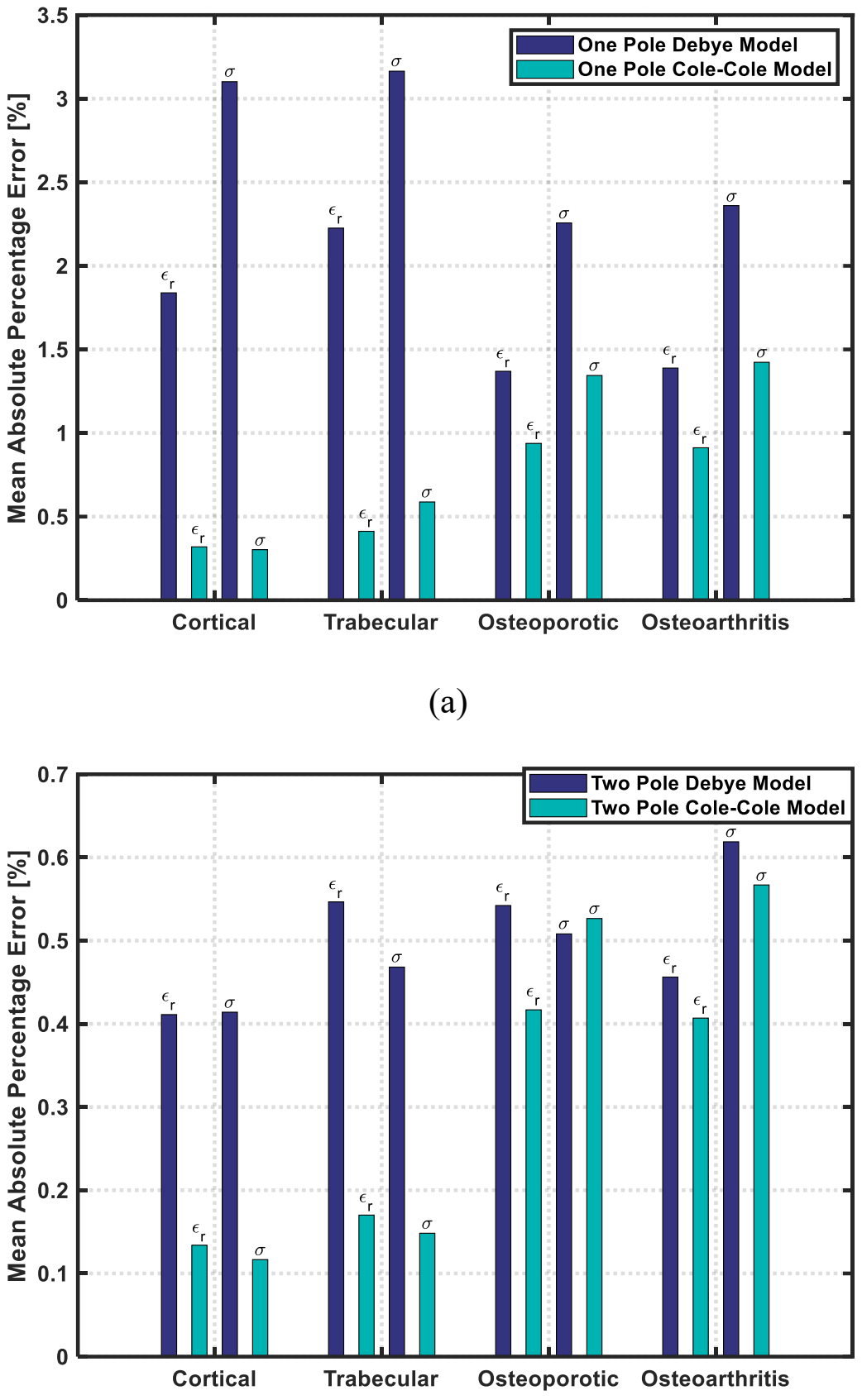

(a)

(b)

**FIGURE 9 Mean Absolute Percentage Error (MAPE) comparison between Debye and Cole-Cole models for the dielectric properties of cortical bone and normal and diseased human trabecular bone (a) Comparison between one-pole Debye and one-pole Cole-Cole model (b) Comparison between two-pole Debye and two-pole Cole-Cole model.**

### *E. Comparative Analysis*

This section compares the prediction accuracy of Cole-Cole and Debye models in terms of Mean Absolute Percentage Error (MAPE). Fig. 9 (a) and (b) show the comparison between the one-pole Cole-Cole and one-pole Debye and two-pole Cole-Cole and two-pole Debye models, respectively. In Fig.9 (a), the one-pole Cole-Cole model shows lower values of MAPE, in comparison to the one-pole Debye model, for the dielectric properties of cortical bone and normal and diseased human trabecular bone. This indicates that the one-pole Cole-Cole model provides a better fit as compared to the one-pole Debye Model. Fig. 9 (b) compares the two-pole Debye and two-pole Cole-Cole models. It can be observed that the two-pole Cole-Cole model shows either less or equal MAPE in comparison to the two-pole Debye model, for the dielectric properties of cortical bone and normal and diseased human trabecular bone. Overall, the two-pole Cole-Cole model provides a better fit as compared to the two-pole Debye Model.

Although the presented models fit the measured data with high accuracy, the system uncertainty itself is 0.61% for relative permittivity and 2.54% for conductivity measurements [28]. Furthermore, for utilising the presented models in studying the dielectric contrast between healthy and diseased trabecular bones, it should be noted that the measured dielectric properties of diseased trabecular bones from [28] and normal trabecular bone from [32] have been taken from different equipment and at different times, which can affect the accuracy of the contrast between the two.

Furthermore, in this study, the dielectric models for cortical and healthy trabecular bones were fitted to the dielectric data estimated from Gabriel's parametric models [32] for completeness. To quantify the consistency and potential error introduced by this approach, we calculated the RMSEs of Gabriel's models for cortical and normal trabecular bone against the original raw measured data reported in [40]. The cortical bone model estimates the relative permittivity and conductivity with RMSEs of 0.26 and 0.08 S/m, respectively, over the frequency range of 0.5 – 8.5 GHz, while the trabecular bone model yields RMSEs of 1.33 and 0.207 S/m, respectively. Furthermore, the RMSE values of our fitted dielectric models relative to Gabriel's models are presented in Tables 2, 4, 6, and 8. The maximum average deviations between our fitted models and Gabriel's model are only 0.45 and 0.06 S/m for trabecular bone, and 0.21 and 0.03 S/m for cortical bone, in terms of relative permittivity and conductivity, respectively. The frequency dependent error plots in Figures 2, 4, 6, and 8 further illustrate that the additional error introduced by re-fitting is minimal and within acceptable limits.

## IV. CONCLUSION

This paper presents one-pole and two-pole Cole-Cole and Debye models for normal and diseased human bone tissue dielectric properties over the microwave frequency range of 0.5 – 8.5 GHz. The models were developed based on the experimental characterisation of the dielectric properties of human trabecular bone samples obtained from osteoporotic and osteoarthritis patients reported by the authors, as well as the dielectric properties of normal trabecular and cortical bone from the literature. The models' parameters were optimised by employing the W-LSM, and the accuracy of the models was evaluated using RMSE and MAPE. Our findings show that the one-pole Cole-Cole model provides a very good fit to the experimental data, with low RMSE and MAPE values for all bone types, indicating the high accuracy of the model compared to the one-pole Debye model. Moreover, the two-pole Cole-Cole model provides a very good fit to the experimental data, with low RMSE and MAPE values for all the bone types compared to the two-pole Debye model. These highly accurate dielectric models can be utilised in complex numerical simulations for the computational investigation of EM wave propagation within the human body, which are vital in dosimetry calculations and the development of EM medical devices. While frequently reported and used for accurate EM modelling, these dielectric models are not without their limitations, including assumptions made in their formulation and potential simplifications of complex biological systems. Despite these limitations, these models serve as valuable tools for understanding bone dielectric properties and have the potential to guide future research endeavours aimed at refining computational models and advancing clinical applications. The Cole-Cole and Debye models are particularly useful for applications like microwave bone imaging and sensor-enabled bone implants. By accurately modelling the distribution of EM energy within bone tissue, simulations can optimise system parameters to enhance imaging, sensing, or device performance. Furthermore, these models improve the fidelity of biomechanical simulations and facilitate the development of sensor-enabled implants and other biomedical devices. Future studies should investigate and compare different optimisation algorithms beyond the weighted least squares, including more advanced techniques such as the two-stage Genetic Algorithm (GA). This would help identify the most suitable optimisation method for accurately fitting the dielectric models to experimental data.

## ACKNOWLEDGMENT

The research leading to these results has received funding from the Future*Care* Seed Fund 2023 by the College of Medicine, Nursing and Health Sciences, University of Galway, Ireland and in part within the research project "3BAtwin - Bone, Brain, Breast and Axillary Medical Microwave Imaging Twinning" funded by Horizon Europe Framework Programme (101159623).

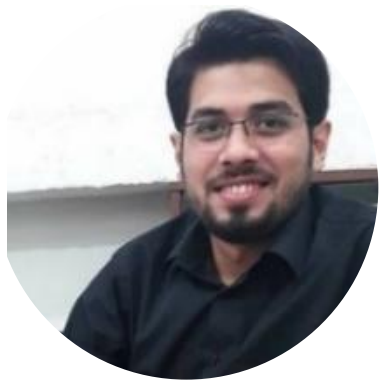

**Dr Bilal Amin** is a Postdoctoral Researcher at the School of Engineering, University of Galway, Ireland. He received his PhD in Electrical and Electronics Engineering from the University of Galway, Ireland. He did his BS in 2013, securing First Class, in Electrical Engineering from COMSATS University Lahore, Pakistan under the auspices of the National ICT scholarship program. In 2015, he earned his master's degree with distinction, in Electrical Engineering from COMSATS University Islamabad, Pakistan. Dr Bilal's main research interest is bioelectromagnetic in healthcare, with particular attention to microwave bone imaging, dielectric characterization of biological tissues, cardiac implant antenna design, microwave ablation antenna design, and medical signal processing with an academic background in clinical and biomedical engineering. He has received over 12 national and international awards and is a member of COST (European Cooperation in Science and Technology) Action MiMED and COST Action EMF-MED.

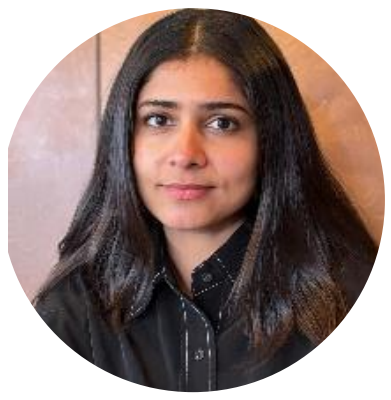

**Amra Mehboob** is a PhD candidate in Electrical and Electronic Engineering at the University of Galway, Ireland, affiliated with the Translational Medical Device Lab. She received her B.S. degree with honours in Telecommunication Engineering from the University of Engineering and Technology (UET), Taxila, Pakistan, in 2015, and an M.S. in Electronic Engineering from the Ghulam Ishaq Khan Institute of Engineering Sciences and Technology (GIKI), Pakistan, in 2018, under a fully funded scholarship. Before her doctoral studies, she worked as an R&D Engineer specializing in antenna design for radar applications. Her current research, supported by the EU-funded 3BAtwin project, focuses on the development of microwave imaging techniques for bone health monitoring. Her research interests include microwave imaging, bioelectromagnetics, dielectric characterization, antenna design, and biomedical innovation.

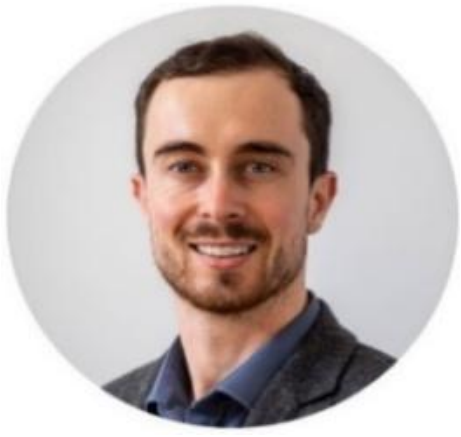

**Dr Eoghan Dunne** (Member, IEEE) is a Post-Doctoral Researcher in the School of Medicine at the University of Galway and is a part of the Translational Medical Device Lab. He received a BEng in Electronic & Computer Engineering (2016) and a PhD in Electrical and Electronic Engineering (2021) from the University of Galway. His Ph. D was focused on using Machine Learning and Electrical Impedance Tomography for monitoring bladder fullness. Since then, he has been working on a Government of Ireland Disruptive Technologies Innovation Fund (DTIF) funded project to help design a pulsed field ablation device to treat Atrial Fibrillation with AuriGen Medical. In the past, he has contributed to international projects including OpenWorm c302 and the European-funded project Si Elegans. His research interests relate to electrical impedance, dielectric and thermal properties as well as the application of electromagnetic-based ablation and electrical impedance imaging.

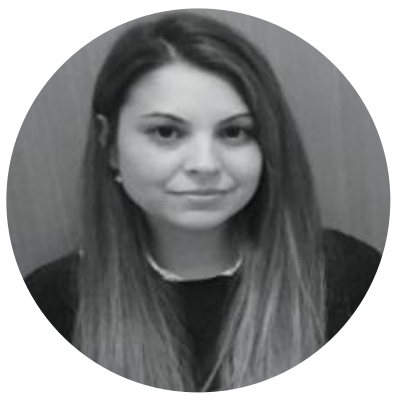

**Dr Ana González-Suárez** is a Research Fellow and has an Honorary Adjunct Lectureship Appointment at the University of Galway, Ireland. She holds a PhD in Biomedical Engineering from the Universitat Politècnica de València, Spain. Her main interests are focused on energy-based ablative techniques, especially in the field of pulsed field ablation of arrhythmias and cancer. She is currently leading the SFI-IRC Pathway project "HEART-SAVE" (22/PATH-S/10719) focused on developing a non-invasive disruptive medical technology to precisely non-thermally ablate the source-site of arrhythmias. Her entire research career has been funded by competitive predoctoral and postdoctoral grants. She has personal secured in direct funding €1.5 million and has published more than 45 papers in peer-review journals. She also received the IFMBE Young Investigator Award and several educational awards. She is currently a commission member of the Woman & Mathematics Committee of the Spanish Royal Mathematical Society and member of a European COST Action.

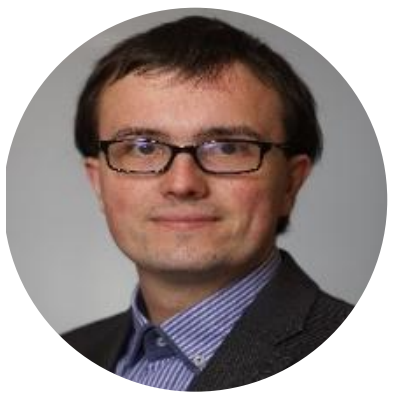

**Dr Finn Krewer** is a Research Fellow at the University of Galway, Ireland. He holds a PhD in Electronic and Computer Engineering from the Bio-Inspired Reconfigurable Computing lab at the University of Galway, Ireland. His work focuses on machine learning and artificial intelligence algorithms in medical and healthcare applications. He has published multiple peer-reviewed journal articles on the topics of tissue dielectric modelling and simulations.

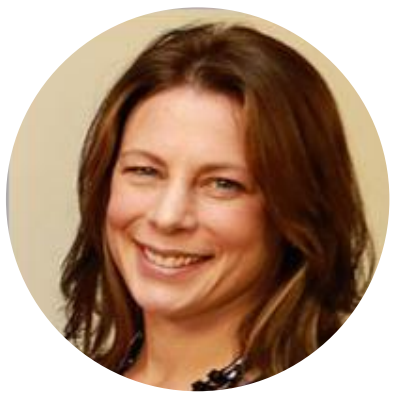

**Prof. Aoife Lowery** is a consultant breast and endocrine surgeon at Galway University Hospital and a Professor in Surgery at the University of Galway. A medical graduate of the University of Galway, Aoife completed her specialist surgical training on the RCSI Higher Surgical Training Programme and an International surgical fellowship in Endocrine Surgery AP-HM (Hopitaux Universitaires de Marseille) France. She was one of the first Clinician Scientist Fellows funded by Molecular Medicine Ireland and completed her PhD investigating microRNA expression and function in breast cancer at the University of Galway. Aoife is currently a recipient of the Irish Cancer Society Clinical Research Leadership award and a principal investigator in the Women's Health Initiative cancer survivorship programme co-funded by the Irish Cancer Society and the National Breast Cancer Research Institute. She is an associate director of the Irish Clinical Academic Training (ICAT) PhD programme a clinical advisor and mentor on the Bioinnovate Ireland fellowship programme and a Clinical Lead for the Clinical Research Facility at the University of Galway.

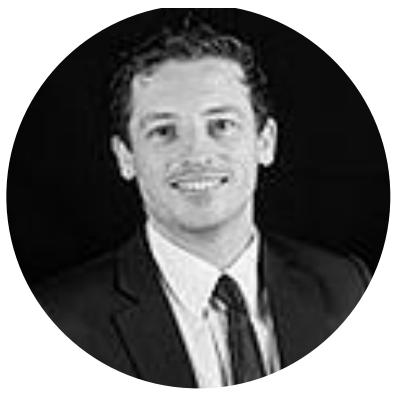

**Prof. Martin O'Halloran** received a B.E. degree in Electronic and Computer Engineering from the University of Galway in 2004, and a Ph.D. in Electrical Engineering also from the University of Galway in 2009. He is currently a full Professor of Medical Electronics at the University of Galway, with a research focus on applied medical device development. In parallel, he leads Ireland's BioInnovate programme, a sister programme of Stanford's BioDesign. Finally, he is a 7-time recipient of funding from the European Research Council (ERC).

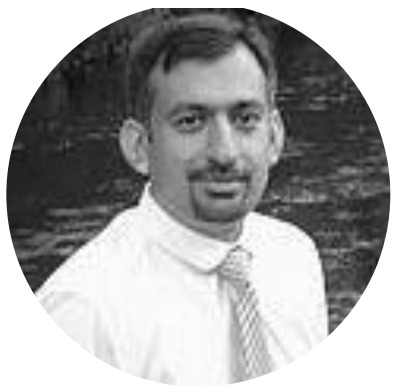

**Dr Adnan Elahi** (Member, IEEE) is a lecturer in medical electronics at the University of Galway, Ireland. He holds a PhD in Electronic Engineering from the University of Galway, an M.Sc. in Embedded Digital Systems from the University of Sussex, United Kingdom and a BS in Computer Engineering from COMSATS University, Lahore, Pakistan. His PhD research was focused on the investigation and development of novel signal processing algorithms to improve microwave imaging of the breast. He has over 8 years of research experience in medical device development. He has previously worked as a lecturer at COMSATS University Lahore, Pakistan; and as a Research Associate at Computer Vision Research Group (COMVIS), Lahore, Pakistan.